\documentclass[reqno,11pt]{amsproc}
\usepackage{amssymb}
\usepackage{amsmath}
\usepackage{amsfonts}
\usepackage{microtype}
\usepackage{xcolor}
\usepackage{geometry}
\usepackage{mathrsfs}
\usepackage{mathtools}

\usepackage[backend=biber,style=ieee,sorting=none,backref=true,citestyle=numeric-comp,giveninits=false]{biblatex}
\usepackage{tikz,tikz-cd}
\usetikzlibrary{arrows,cd,decorations.markings,arrows.meta}

\tikzset{->-/.style={decoration={markings,mark=at position .5 with {\arrow{>}}},postaction={decorate}}}
\tikzset{-<-/.style={decoration={markings,mark=at position .5 with {\arrow{<}}},postaction={decorate}}}
\tikzstyle{box}=[fill=white, draw=black, shape=rectangle, inner sep=14pt]
\tikzstyle{forward arrow}=[->-]
\tikzstyle{backward arrow}=[-<-]

\definecolor{myurlcolor}{rgb}{0,0,0.4}
\definecolor{mycitecolor}{rgb}{0,0.5,0}
\definecolor{myrefcolor}{rgb}{0.5,0,0}
\usepackage[breaklinks=true]{hyperref}
\hypersetup{colorlinks,
linkcolor=myrefcolor,
citecolor=mycitecolor,
urlcolor=myurlcolor}
\usepackage[capitalize]{cleveref}

\newcommand{\beq}{\begin{equation}}
\newcommand{\eeq}{\end{equation}}

\newcommand{\R}{\mathbb{R}}
\newcommand{\C}{\mathbb{C}}

\theoremstyle{plain}
\newtheorem{dummy}{Dummy}[section]
\Crefname{thm}{Theorem}{Theorems}
\Crefname{lem}{Lemma}{Lemmas}
\Crefname{prop}{Proposition}{Propositions}
\Crefname{cor}{Corollary}{Corollaries}
\Crefname{ass}{Assumption}{Assumptions}
\Crefname{qstn}{Question}{Questions}
\newtheorem{defn}[dummy]{Definition}\Crefname{defn}{Definition}{Definitions}
\Crefname{nota}{Notation}{Notations}
\Crefname{prob}{Problem}{Problems}
\Crefname{slog}{Slogan}{Slogans}
\theoremstyle{remark}
\newtheorem{ex}[dummy]{Example}\Crefname{ex}{Example}{Examples}
\Crefname{conj}{Conjecture}{Conjectures}
\Crefname{rem}{Remark}{Remarks}
\Crefname{note}{Note}{Notes}
\numberwithin{equation}{section}
\Crefname{equation}{}{}		

\Crefname{scalar_ex}{Scalar Field}{Scalar Field}
\Crefname{electro_ex}{Electrodynamic Field}{Electrodynamic Field}
\Crefname{spinor_electro_ex}{Spinor Field}{Spinor Field}
\Crefname{spinor_electro_ex}{Spinor Electrodynamics}{Spinor Electrodynamics}
\Crefname{gr_ex}{Metric Field}{Metric Field}
\Crefname{scalar_fun_ex}{Scalar Field (Kinematic)}{Scalar Field (Kinematic)}

\allowdisplaybreaks

\let\originalleft\left
\let\originalright\right
\renewcommand{\left}{\mathopen{}\mathclose\bgroup\originalleft}
\renewcommand{\right}{\aftergroup\egroup\originalright}

\usepackage{enumitem}
\setlist[enumerate]{label=(\roman*),itemsep=5pt,topsep=8pt}
\setlist[itemize]{label=$\triangleright$,itemsep=5pt,topsep=6pt}
\Crefformat{enumi}{#2#1#3}

\newcommand{\newterm}[1]{\textbf{#1}}

\usepackage{hyphenat}
\begin{document}



\title{Self-distributive structures in physics}

\author{Tobias Fritz}

\address{Department of Mathematics, University of Innsbruck, Austria}
\email{tobias.fritz@uibk.ac.at}



\thanks{\textit{Acknowledgements.} We thank Benjamin Walder for helpful discussions on Lie quandles, John Baez and Markus Szymik for further discussion and valuable pointers to the literature and the anonymous referee for suggesting interesting examples of Noether quandles.}

\begin{abstract}
	It is an important feature of our existing physical theories that observables generate one-parameter groups of transformations.
	In classical Hamiltonian mechanics and quantum mechanics, this is due to the fact that the observables form a Lie algebra, and it manifests itself in Noether's theorem.
	In this paper, we propose \emph{Lie quandles} as the minimal mathematical structure needed to express the idea that observables generate transformations.
	This is based on the notion of a quandle used most famously in knot theory, whose main defining property is the self-distributivity equation $x \triangleright (y \triangleright z) = (x \triangleright y) \triangleright (x \triangleright z)$.
	We argue that Lie quandles can be thought of as nonlinear generalizations of Lie algebras.

	We also observe that taking convex combinations of points in vector spaces, which physically corresponds to mixing states, satisfies the same form of self-distributivity.
\end{abstract}

\maketitle

\section{Introduction}
\label{introduction}

Observables play a dual role in our existing physical theories~\cite{baez}.
On the one hand, they are the quantities that we think of as being measurable in principle.
On the other hand, they are also exactly those quantities that generate groups of transformations, including symmetries and the dynamics of a theory.
This surprising dual role is especially apparent and well-understood in Hamiltonian mechanics, where it is closely tied to the symplectic structure of phase space, and in quantum mechanics, where it manifests itself in the correspondence between self-adjoint operators and one-parameter groups of unitaries (Stone's theorem).
In both cases, this duality is what underlies Noether's theorem: an observable is a conserved quantity if and only if the transformation group that it generates is a group of symmetries.

This fact that observables generate transformations is not an a priori necessary feature of physical theories.
In fact, it is not difficult to come up with candidate physical theories which violate it.
For example, this is generically the case for the \emph{general probabilistic theories} studied in quantum foundations~\cite{barrett,plavala}: in a general probabilistic theory, a state space is given by a convex body in $\R^n$, and a generic such object has no symmetries at all, and in particular no nontrivial one-parameter groups of transformation.
Another class of theories in which observables cannot generate transformations in the way that we are used to are theories with non-reversible time evolution.
What we are used to is that if dynamics is generated by an observable denoted $H$, then we would expect there to be another observable $-H$ which generates the time reversed dynamics.

We are thus led to conclude that having a well-behaved correspondence between observables and one-parameter groups of transformations is a distinguishing feature of those theories that describe the physics of our universe (as far as we understand it).
Suppose now that we want to develop other theories---perhaps more fundamental ones---that follow this very principle. Then which mathematical structure do we need in order to capture this feature?
In Hamiltonian mechanics and quantum mechanics, this principle is encoded in the fact that the spaces of observables is a \newterm{Lie algebra}.
This leads to the desired feature because any observable $H$ generates a one-parameter group of transformations \emph{on the very space of observables itself} determined by the equation
\begin{equation}
	\label{heisenberg_evolution}
	\dot{A} = [H, A].
\end{equation}
In quantum mechanics, the Lie bracket is exactly the usual commutator (up to factors of $i$ and $\hbar$ which we ignore), so that this specializes to the usual time evolution equation in the Heisenberg picture.
In Hamiltonian mechanics, the Lie bracket is the Poisson bracket, so that this equation Hamilton's equations of motion in their general abstract form.
In both cases, our notation suggests thinking of $H$ as the Hamiltonian and of the transformations as time evolution.
But of course it applies likewise if $H$ is e.g.~the momentum operator instead, in which case the transformed observable $A(t)$ is a spatially translated version of $A$.

What we argue in this paper is that the Lie algebra structure is overkill in order to formalize the idea of a well-behaved map from observables to transformation groups.
Instead, we posit that it is enough to have a \newterm{Lie quandle}, which is an algebraic structure that we introduce here and think of as a nonlinear generalization of a Lie algebra.
To motivate it, let us simply axiomatize the idea that observables generate transformations on the space of observables itself a standalone algebraic structure. So given an observable $H$ and another observable $A$, for every $t \in \R$ we should get a new observable
\begin{equation}
	\label{observable_action}
	H \triangleright_t A,
\end{equation}
where this notation is to be read as `$H$ acting on $A$ for time $t$'.
Leaving $A$ unspecified means that we have a map $H \triangleright_t -$ from the set of observables to itself.
Since this should be a one-parameter group of transformations, we expect this to be additive in the time argument: we should have the equations
\begin{equation}
	\label{additivity}
	H \triangleright_{t + s} A = H \triangleright_t (H \triangleright_s A), \qquad H \triangleright_0 A = A.
\end{equation}
Furthermore, it is natural to expect that the maps $H \triangleright_t -$ respect whatever mathematical structure the set of observables has.
But surely \emph{the binary operations $\triangleright_s$ themselves} are part of this structure.
Hence we expect every $H \triangleright_t -$ to be a map which is a homomorphism with respect to every $\triangleright_s$.
This means that for all observables $H$, $A$ and $B$, we should have
\begin{equation}
	\label{self_distributivity0}
	H \triangleright_t (A \triangleright_s B) = (H \triangleright_t A) \triangleright_s (H \triangleright_t B) \qquad \forall s,t \in \R.
\end{equation}
This intriguingly self-referential equation is a form of \newterm{self-distributivity}, which famously appears in the theory of self-distributive structures like \emph{quandles}~\cite{joyce,EN}.
Surprisingly, the role of such self-distributivity in physics does not seem to have been appreciated before\footnote{One notable exception is a blog post by Daniel Moskovich~\cite{moskovich}, which expresses the idea that associative algebraic structures are geometric in nature, while self-distributive operations are appropriate for `information physics'. However, he does not seem to have recognized the role of self-distributivity in the correspondence between observables and transformation groups.}, and quandles have hitherto mainly been considered in pure mathematics\footnote{Their most well-known application is to knot theory~\cite{burde,joyce,matveev}.
	However, the earliest appearance of (involutive) quandles seems to be as algebraic structures abstracting reflections~\cite{takasaki}, and they have also been used in number theory~\cite{LL} and real algebra~\cite{szymik}. Brieskorn's early survey~\cite{brieskorn} discusses relations to braid groups, root systems and singularity theory.}.
Hence the main purpose of this paper is to point out the relevance of self-distributivity in physics.

There is another appearance of self-distributivity in physics which underlines its relevance, but now at the level of \emph{states} rather than observables.
In quantum theory, forming convex combinations or \newterm{mixtures} of (mixed) states has been recognized as a crucial piece of structure already by von Neumann~\cite[Ch.~IV]{vonneumann}, and has since become central through the framework of general probabilistic theories.
Now if we consider convex combinations operations in the form
\[
	x \ast_t y \coloneqq (1 - e^{-t}) x + e^{-t} y,
\]
then this family of binary operations \emph{again} turns out to the satisfy the properties~\eqref{additivity} and~\eqref{self_distributivity0}, which now take the form
\begin{align*}
	x \ast_{t + s} y = x \ast_t (x \ast_s y), \qquad & \quad x \ast_0 y = y, \\[4pt]
	x \ast_t (y \ast_s z) = (x \ast_t y) & \ast_s (x \ast_t z).
\end{align*}
These equations are straightforward to show by direct calculation.
In light of our earlier discussion, one may think of this as saying that an operation of the form $x \ast_t -$ is a kind of dynamics on the set of states, which `shrinks' the whole state space in the direction of $x$.
The self-distributivity equation may be thought of as saying that this operation preserves \emph{other} convex combinations; this is clear even without calculation based on the fact that every affine map preserves convex combinations, and the map $x \ast_t -$ is affine.

It remains mysterious to us whether the double appearance of self-distributivity at the level of observables and at the level of states is somehow coincidental, or whether there is a deeper underlying connection between the two.
A point in favour of a deeper connection is that both $\triangleright_t$ and $\ast_t$ satisfy one further equation which we have not mentioned yet, namely idempotency:
\[
	A \triangleright_t A = A, \qquad x \ast_t x = x.
\]
On the other hand, there also is an important difference between the two cases, which suggests that the similarity may be coincidental.
To wit, forming convex combinations also satisfies \emph{right} self-distributivity,
\[
	(x \ast_t y) \ast_s z = (x \ast_t z) \ast_s (y \ast_t z),
\]
which is an equation that typically does not hold for the $\triangleright_t$ operations.

\subsection*{Related Work}

While our notion of Lie quandle and its physical interpretation in terms of observables generating transformations seem to be new, the importance of self-distributivity in physics has been recognized in a blog post by Moskovich~\cite{moskovich}.
There also are a number of existing works relating self-distributive structures with Lie theory:
\begin{itemize}
	\item 
The thesis of Crans~\cite[Section~3.2.4]{crans} considers unital spindles and quandles \emph{internal} to the category of coalgebras, and uses these structures as an intermediate notion between Lie groups and Lie algebras.
Since the underlying set of Crans' quandles correspond to the \emph{symmetric powers} of Lie algebras, this seems quite different from our Lie quandles (of which the simplest examples are Lie algebras themselves).
\item \emph{Smooth quandles} were introduced by Ishikawa~\cite{ishikawa}\footnote{There is a more general notion of quandles in categories with finite products~\cite{grosfjeld}, which yields smooth quandles upon taking the category to be the category of manifolds.}; we also refer to the thesis of Yonemura~\cite{yonemura} for a detailed exposition.
		Smooth quandles are quandles internal to the category of manifolds (meaning that the underlying set is a manifold and the quandle operation is smooth).
		This is more closely related to Lie quandles: a Lie quandle becomes a smooth quandle if one forgets all operations $\triangleright_t$ for $t \neq 1$.

		Nevertheless, having all $\triangleright_t$ available to work with seems more prudent for our purposes: it reflects the physical structure more accurately, and it is also mathematically convenient since we can differentiate as e.g.~at~\eqref{lie_bracket}.
\end{itemize}

\section{Self-distributivity}

To begin a more detailed discussion, let us first consider the mathematics of self-distributivity in its essence.
We will not do anything original here, and we refer to literature like~\cite{dehornoy,joyce,crans,EN} for more details.

\begin{defn}[{e.g.~\cite{dehornoy,crans}}]
	\label{shelf}
	A \newterm{shelf} is a set $S$ together with a binary operation $\triangleright$ such that
	\begin{equation}
		\label{self-distributivity2}
		x \triangleright (y \triangleright z) = (x \triangleright y) \triangleright (x \triangleright z)
	\end{equation}
	for all $x, y, z \in S$.
\end{defn}

Here we have adopted the terminology of Crans~\cite[Definition~52]{crans}\footnote{We omit the attribute `left' since we will not consider right shelves.}.
A good way to think about this equation is as saying that the map $x \triangleright - : S \to S$ is a homomorphism with respect to $\triangleright$ itself for every $x \in S$.
In this way, shelves axiomatize the idea of a set acting on `itshelf'.

\begin{ex}
	The set $S = \{0,1,2\}$ with the binary operation
	\begin{equation}
		\label{cyclic_shelf}
		\begin{tabular}{c|ccc}
  			$\triangleright$ & 0 & 1 & 2 \\
	  		\hline
  			0 & 0 & 2 & 1 \\
  			1 & 2 & 1 & 0 \\
  			2 & 1 & 0 & 2 \\
		\end{tabular}	
	\end{equation}
	may be the historically first example of a shelf~\cite[p.~481]{dehornoy}.
	Note that every element acts by permuting the other two.
\end{ex}

Of particular interest to us are shelves which also satisfy the idempotency equation.

\begin{defn}[{\cite{dehornoy,crans}}]
	A \newterm{spindle} is a shelf $S$ such that
	\[
		x \triangleright x = x
	\]
	for all $x \in S$.
\end{defn}

\begin{ex}
	\label{spindle_examples}
	\begin{enumerate}
		\item A glance at the diagonal entries of~\eqref{cyclic_shelf} shows that that shelf is indeed a spindle.
		\item\label{reflection_rotation} For another example, take $S \coloneqq \R^2$ and define $x \triangleright y$ to be the point obtained by point reflecting $y$ across $x$.
			This satisfies self-distributivity due to the fact that \emph{any} Euclidean transformation preserves distances, the point $x \triangleright y$ can be characterized uniquely in terms of its distances from $x$ and $y$,\footnote{Namely as the unique point whose distance from $x$ equals the distance of $x$ and $y$, and whose distance from $y$ is twice that.} and reflection at $x$ is a Euclidean transformation.
		\item More generally, we can take $S$ to be any \emph{symmetric space}, such as the sphere $S^2$, and define $x \triangleright y$ to be the point obtained by reflecting $y$ across $x$~\cite[Example~56]{EN}\footnote{Note that~\cite{EN} considers \emph{right} self-distributivity rather than the left version of~\Cref{shelf}, so that their $x \triangleright y$ corresponds to our $y \triangleright x$.}.
		\item For any fixed $\theta \in [0,2\pi)$, we get another spindle structure on $\R^2$ by defining $x \triangleright y$ to be the point obtained by \emph{rotating} $y$ around $x$ by $\theta$; taking $\theta = \pi$ then recovers the previous example.
		\item Similarly, we can work with points on the sphere $S^2$ and define $x \triangleright y$ to be the point obtained by rotating $y$ around $x$ by $\theta$~\cite[Definition~4.3]{CS}.
	\end{enumerate}
\end{ex}

\begin{ex}
	The following example is a simple variation on the standard \emph{Alexander quandle} from the theory of knot invariants~\cite[Example~67]{EN}.
	It also appears in Moskovich's blog post~\cite{moskovich} and has probably been known before.

	Let $C \subseteq \R^n$ be a convex set (such as the set of mixed states of a physical system).
	Then for any $s \in [0,1]$, the set $C$ is a spindle with respect to the operation of mixing with bias $s$,
	\begin{equation}
		\label{convex_combination}
		x \triangleright y \coloneqq (1-s)x + sy.
	\end{equation}
	Intuitively, the self-distributivity equation~\eqref{self-distributivity2} holds because taking a convex combination with $x$ uniformly shrinks all points in the direction of $x$, and this operation preserves other convex combinations.
\end{ex}

The following particularly elegant types of spindles can be thought of as the analogues of groups in the self-distributive world.

\begin{defn}
	A \newterm{quandle} $Q$ is a spindle such that for every $x \in Q$, the map
	\[
		x \triangleright -: Q \to Q
	\]
	is a bijection.
\end{defn}

If we denote the inverse map by $x \triangleright_{-1} - : Q \to Q$, and moreover write $\triangleright_1$ in place of $\triangleright$, then the fact that these maps are inverse amounts to the equation
\begin{equation}
	\label{quandle_inverse}
	x \triangleright_{-1} (x \triangleright_1 y) = y, \qquad x \triangleright_1 (x \triangleright_{-1} y) = y.
\end{equation}
Moreover, it is easily seen that $\triangleright_{-1}$ equips the original set $Q$ with another quandle structure,\footnote{Some authors use the opposite notational convention by introducing an operation $\triangleleft$ given by $x \triangleleft y \coloneqq y \triangleright_{-1} x$, which then satisfies the \emph{right} self-distributivity equation $(x \triangleleft y) \triangleleft z = (x \triangleleft z) \triangleleft (y \triangleleft z)$~\cite[Section~3.1.1]{crans}.} and $\triangleright_1$ and $\triangleright_{-1}$ also mutually distribute over each other~\cite[p.~95]{EN}.
These statements are observations that naturally lead to the Lie quandles of \Cref{lie_quandle} if one considers an entire family of operations $\triangleright_t$ indexed by $t \in \R$.

\begin{ex}
	\label{conjugation_quandle}
	\begin{enumerate}
		\item Assuming $s \in (0,1)$, a convex set $C \subseteq \R^n$ is typically not a quandle with respect to~\eqref{convex_combination}, since for given $x, z \in C$ the point $y$ with $(1 - s)x + sy = z$ need not belong to $C$ again.
			In fact, it is easy to see that such $C$ is a quandle if and only if it is an affine subspace of $\R^n$.
		\item For a group $G$, its \newterm{conjugation quandle} is $Q \coloneqq G$ itself equipped with conjugation as the operation,
			\[
				x \triangleright y \coloneqq xyx^{-1}.
			\]
			This is the most important family of examples of quandles.
			These quandles satisfy the following special property: for every given $x, y \in Q$, we have
			\begin{equation}
				\label{prenoether}
				x \triangleright y = y \qquad \Longleftrightarrow \qquad y \triangleright x = x,
			\end{equation}
			We will later find this to be closely related to Noether's theorem.
		\item\label{action_quandle}
			To see that not every quandle satisfies~\eqref{prenoether}, let $X$ be a set on which $G$ acts, and consider the disjoint union $Q \coloneqq G \sqcup X$.
			We define
			\[
				x \triangleright y \coloneqq
				\begin{cases}
					x y x^{-1} \in G & \textrm{if } x, y \in G,\\
					x y \in X & \textrm{if } x \in G, \:  y \in X,\\
					y \in X & \textrm{if } x \in X.
				\end{cases}
			\]
			In other words, for $x \in G$ we let $x \triangleright -$ act by conjugation on $G$ and by the given $G$-action on $X$, and for $x \in X$ we take $x \triangleright -$ to be the identity map.
			Then simply checking all cases shows that this is a quandle, but~\eqref{prenoether} is generically not satisfied if $x \in X$ and $y \in G$.

			This construction works the same way if $G$ is replaced by any quandle and the action on $X$ by a \emph{quandle action} in the sense of~\cite[\S~3.2]{elhamdadi}.
	\end{enumerate}
\end{ex}

Hitherto, the theory of quandles has been primarily applied to knot theory.
We refer to the original papers~\cite{burde,joyce,matveev} as well as to the more recent textbook~\cite{EN} for expositions of this application.\footnote{A reader who knows about anyon physics may wonder whether anyons provide another way in which quandles appear in physics. As far as we can see this is not the case, since anyons are modelled by braided fusion categories, and these do not seem to bear any direct relation (with physical significance) to quandles.}

\section{Lie quandles}
\label{lie_quandle}

As indicated in the introduction, we now introduce a Lie-theoretic version of quandles.
From our point of view, these are the minimal structures needed to formalize the idea that observables of a physical theory form a set such that every elements acts on the set itself as a one-parameter group of transformations.

\begin{defn}
	A \newterm{Lie quandle} is a smooth manifold $Q$ together with an operation
	\[
		\triangleright \: : \: Q \times \R \times Q \longrightarrow Q
	\]
	such that the following equations hold for all $x,y,z \in Q$ and $s,t \in \R$:
	\begin{enumerate}
		\item Self-action:
			\begin{equation}
				x \triangleright_s (x \triangleright_t y) = x \triangleright_{s+t} y, \qquad
				x \triangleright_0 y = y.
			\end{equation}
		\item Self-distributivity:
			\begin{equation}
				\label{self-distributivity}
				x \triangleright_s (y \triangleright_t z) = (x \triangleright_s y) \triangleright_t (x \triangleright_s z).
			\end{equation}
		\item Idempotency:
			\begin{equation}
				x \triangleright_s x = x.
			\end{equation}
	\end{enumerate}
\end{defn}

\begin{ex}
	In quantum mechanics, the space of observables is the space of self-adjoint elements of a C*-algebra; let us focus on the finite-dimensional case of a matrix algebra for simplicity.\footnote{In order for our statements to apply to infinite-dimensional C*-algebras, one would need to work with a suitable notion of infinite-dimensional manifold in the definition of a Lie quandle.}
	Then the set of observables is given by the set of hermitian matrices,
	\[
		Q \coloneqq \{ A \in M_n(\C) \mid A^\dag = A \}.
	\]
	This is the paradigmatic example of a Lie quandle with respect to the operations
	\[
		X \triangleright_t Y \coloneqq e^{itX} Y e^{-itX},
	\]
	which is the standard way in which an observable generates a one-parameter group of transformations of other observables in quantum mechanics.
	While the other Lie quandle axioms are quite straightforward to check, let us explain how to verify the self-distributivity equation~\eqref{self-distributivity} explicitly.
	This uses the well-known fact that matrix exponentiation preserves conjugation, that is
	\[
		e^{i t A Y A^{-1}} = A e^{i t Y} A^{-1}
	\]
	for any matrix $Y$ and any invertible $A$~\cite[Proposition~2.3(6)]{hall}.
	We apply this with $A = e^{isX}$ in the third step of the following sequence of equations:
	\begin{align*}
		X \triangleright_s (Y \triangleright_t Z) &= e^{isX} \left( e^{itY} Z e^{-itY} \right) e^{-isX} \\
			&= \left( e^{isX} e^{itY} e^{-isX} \right) \left( e^{isX} Z e^{-isX} \right) \left( e^{isX} e^{-itY} e^{-isX} \right) \\
			&= e^{it\, e^{isX} Y e^{-isX}} \left( e^{isX} Z e^{-isX} \right) e^{it\, e^{isX} Y e^{-isX}} \\
			&= (X \triangleright_s Y) \triangleright_t (X \triangleright_s Z).
	\end{align*}
	This establishes the required self-distributivity.
	While this calculation may seem obscure, the intuition behind it makes it quite unsurprising: conjugation by a unitary preserves all algebraic structure on matrices, \emph{including conjugation by a unitary itself}.
	This also makes clear that these Lie quandles are souped-up versions of the conjugation quandles from \Cref{conjugation_quandle}.

Since we have not used the assumption of $X$ being hermitian in the above calculation, the same calculation can be used to show that the space of \emph{all} matrices $M_n(\C)$ is a Lie quandle with respect to the operation
\[
	X \triangleright_t Y \coloneqq e^{tX} Y e^{-tX}.
\]
This Lie quandle structure can also be characterized in terms of the differential equation
\begin{equation}
	\label{lie_evolution}
	\frac{d}{dt} (X \triangleright_t Y) = [X, X \triangleright_t Y]
\end{equation}
with initial condition $X \triangleright_0 Y = Y$.
This is exactly the evolution equation~\eqref{heisenberg_evolution} from Hamiltonian mechanics and quantum mechanics written in different form.
\end{ex}

The characterization~\eqref{lie_evolution} of the Lie quandle structure implies that every (real) Lie algebra $\mathfrak{g} \subseteq M_n(\C)$ is a Lie quandle with respect to the operations $\triangleright_t$ restricted from all matrices to $\mathfrak{g}$.
By Ado's theorem, we can embed every Lie algebra into $M_n(\C)$ for some $n$, and therefore every Lie algebra is a Lie quandle in a canonical way.
If $G$ is any Lie group with Lie algebra $\mathfrak{g}$, then this Lie quandle structure on $\mathfrak{g}$ can also be written in terms of the exponential map $\exp : \mathfrak{g} \to G$ and the \newterm{adjoint action} of $G$ on $\mathfrak{g}$~\cite[\S 4.3.3]{hall} as
\[
	X \triangleright_t Y = \mathrm{Ad}(e^{tX}) Y,
\]
since this solves the differential equation~\eqref{lie_evolution} with initial condition $X \triangleright_0 Y = Y$.
The differential equation also shows that the Lie bracket can be recovered as
\begin{equation}
	\label{lie_bracket}
	[X, Y] = \frac{d}{dt}\Big|_{t=0} (X \triangleright_t Y) .
\end{equation}
In summary, Lie algebras are Lie quandles, and the Lie bracket is encoded in the Lie quandle structure.
This suggests that we can think of a general Lie quandle as a nonlinear generalization of Lie algebras.

But there also are Lie quandles which are not Lie algebras.
When considered in the physics context, the following nonlinear example feels similar to the Bloch sphere with dynamics given by Rabi oscillations.

\begin{ex}
	\label{bloch_quandle}
	We introduce the \newterm{Bloch quandle}, which is a Lie-theoretic version of the sphere quandle appearing in \Cref{spindle_examples}\Cref{reflection_rotation} by considering all angles of rotation at the same time.
	This amounts to taking $Q \coloneqq S^2$, and for given points $x, y \in S^2$, we define $x \triangleright_t y$ as the point obtained by rotating $y$ by an angle $t$ around the axis having $x \in S^2$ as its north pole.\footnote{Whether one rotates in the positive or negative direction is then irrelevant as long as it is done consistently.}

	One way to see that this satisfies the self-distributivity equation is to note that the geometric structure needed to define these rotations is itself preserved under rotations.
	Another way to show that this defines a Lie quandle is to note that there is an embedding of this structure as a Lie subquandle of the hermitian $2 \times 2$-matrices by identifying $S^2$ with the space of projections of rank $1$.
	In physics terms, this is simply the map which assigns to every pure qubit state the corresponding density matrix.

	Clearly this Lie quandle is not isomorphic to a Lie quandle arising from the Lie algebra, since the underlying manifold $S^2$ is not even diffeomorphic to a vector space.
	Also, analogous constructions of Lie quandles defined by rotations are possible if one replaces $S^2$ by the Euclidean plane or the hyperbolic plane.
\end{ex}

In physical terms, we may want to think of examples like this as (nonlinear) spaces of observables, where each observable defines a Heisenberg-picture-type dynamics on the space of observables itself, as already discussed in the introduction.

The correspondence between observables and one-parameter groups of transformations plays a crucial role in Noether's theorem.\footnote{We refer to~\cite[Theorem~5.58]{olver} for a general version of Noether's theorem in classical field theory.}
We can now understand this result by phrasing a formulation of Noether's theorem due to Baez~\cite{baez} in the framework of Lie quandles.

\begin{defn}
	A \newterm{Noether quandle}\footnote{Although calling this a `Noetherian Lie quandle' would be intrinsically more intuitive, the clash with the meaning of `Noetherian' in commutative algebra seems too severe to justify the more intuitive term.} is a Lie quandle $Q$ such that
	\begin{equation}
		\label{noether}
		x \triangleright_t y = y \quad \forall t \in \R \qquad \Longleftrightarrow \qquad y \triangleright_t x = x \quad \forall t \in \R.
	\end{equation}
\end{defn}

Here is how this corresponds to Noether's theorem.
If $y$ plays the role of a Hamiltonian, then the condition $x \triangleright_t y = y$ for all $t \in \R$ means that the observable $x$ generates a group of transformations that are symmetries of the Hamiltonian.
On the other hand, the condition $y \triangleright_t x = x$ can be interpreted as a conservation law: the observable $x$ is invariant under the dynamics generated by the Hamiltonian $y$.
Therefore Noether's theorem is equivalent to~\eqref{noether}.

\begin{ex}[cf.~\cite{baez}]
	The Lie quandle associated to a Lie algebra $\mathfrak{g}$ is a Noether quandle, since in this case the condition $x \triangleright_t y = y$ for all $t \in \R$ is equivalent to $[x, y] = 0$.
	Then~\eqref{noether} amounts to the equivalence between $[x, y] = 0$ and $[y, x] = 0$, which is the antisymmetry of the Lie bracket.
	This is the straightforward proof of Noether's theorem in Hamiltonian mechanics and quantum mechanics.
\end{ex}

\begin{ex}
	The Bloch quandle from \Cref{bloch_quandle} is a Noether quandle.
	One way to see this is to recall that it is a Lie subquandle of a Noether quandle, and the Noether property~\eqref{noether} is preserved by the passage to a Lie subquandle.
	
	The analogously defined Lie quandles based on the Euclidean plane or the hyperbolic plane are Noether quandles as well.
	Indeed in all three cases, the equation $x \triangleright_t y = y$ for all $t \in \R$ is easily seen to be equivalent to $x = y$.
\end{ex}

We owe the next example of a Noether quandles to the referee.

\begin{ex}
	In a similar spirit to the Bloch quandle, fix distinct numbers $\lambda_1, \ldots, \lambda_n \in \R$, and consider the set of all hermitian matrices in $M_n(\C)$ with spectrum equal to $\{ \lambda_1, \ldots, \lambda_n \}$.
	By the spectral theorem, these matrices are in bijection with orthonormal bases of $\C^n$.
	Therefore this set is a submanifold of $M_n(\C)$.
	Since conjugating by a unitary leaves the spectrum invariant, we obtain a Lie subquandle of $M_n(\C)$, which is therefore a Noether quandle as well.
\end{ex}

\begin{ex}
	To see that \emph{not} every Lie quandle is a Noether quandle, we can use a similar construction as in \Cref{conjugation_quandle}\ref{action_quandle}.
	Indeed if a Lie group $G$ acts on a manifold $X$, then the disjoint union $Q \coloneqq \mathfrak{g} \sqcup X$ becomes a Lie quandle with respect to the operations $x \triangleright_t -$ defined such that every $x \in X$ acts trivially, and such that $x \in \mathfrak{g}$ acts on $\mathfrak{g}$ by the adjoint action as above and by the given Lie group action of $e^{tx}$ on $X$.
	Similar to what we saw in \Cref{conjugation_quandle}\ref{action_quandle} with the failure of \eqref{prenoether}, this Lie quandle is not a Noether quandle as soon as \emph{some} exponential of an element of $\mathfrak{g}$ acts nontrivially on \emph{some} element of $X$.
\end{ex}

On the other hand, it is still conceivable that every \emph{connected} Lie quandle is a Noether quandle.
If this does turn out to be the case, then we have a very general form of Noether's theorem.

\section{Conclusion}

We have argued that self-distributive structures deserve to play a prominent role in physics, upending the usual focus on associative structures.
This is in line with Moskovich's idea that associative algebraic structures are geometric in nature, while self-distributive operations are appropriate for `information physics'~\cite{moskovich}.
The basic observation is that observables form a self-distributive structure with respect to every observable generating a one-parameter transformation group on the space of observables, and we have proposed a definition of Lie quandle to formalize this structure more generally, and we have argued that Lie quandles are nonlinear generalizations of Lie algebras.

Of course, a space of observables can also be expected to carry structure that we have not considered here. 
For example, we have not investigated in what way observables have `values' and how these values should be expected to interact with the quandle structure.
Nevertheless, we believe that Lie quandles provide part of an interesting framework for physical theories generalizing Hamiltonian mechanics and quantum mechanics.

\bigskip

\subsection{Funding and/or Conflicts of interests/Competing interests}

No funding was received to assist with the preparation of this manuscript.
The author has no competing interests to declare that are relevant to the content of this article.

\newpage
\printbibliography

\end{document}